# Model-Based Photoacoustic Image Reconstruction using Compressed Sensing and Smoothed L0 Norm

Moein Mozaffarzadeh[a], Ali Mahloojifar[a,*], Mohammadreza Nasiriavanaki[b], and Mahdi Orooji[a]

[a]Department of Biomedical Engineering, Tarbiat Modares University, Tehran, Iran
[b]Department of Biomedical Engineering, Wayne State University, Detroit, Michigan, USA

**ABSTRACT**

Photoacoustic imaging (PAI) is a novel medical imaging modality that uses the advantages of the spatial resolution of ultrasound imaging and the high contrast of pure optical imaging. Analytical algorithms are usually employed to reconstruct the photoacoustic (PA) images as a result of their simple implementation. However, they provide a low accurate image. Model-based (MB) algorithms are used to improve the image quality and accuracy while a large number of transducers and data acquisition are needed. In this paper, we have combined the theory of compressed sensing (CS) with MB algorithms to reduce the number of transducer. Smoothed version of $\ell_0$-norm (S$\ell_0$) was proposed as the reconstruction method, and it was compared with simple iterative reconstruction (IR) and basis pursuit. The results show that S$\ell_0$ provides a higher image quality in comparison with other methods while a low number of transducers were. Quantitative comparison demonstrates that, at the same condition, the S$\ell_0$ leads to a peak-signal-to-noise ratio for about two times of the basis pursuit.

## 1. INTRODUCTION

Photoacoustic imaging (PAI) is a novel medical imaging modality that uses the advantages of the spatial resolution of ultrasound imaging and the high contrast of pure optical imaging.[1–3] The merits of Photoacoustic (PA) effects have been shown in many investigations.[2,4–10] A numerical analysis for infant brain imaging has been conducted.[11] Low-cost photoacoustic imaging systems have been extensively investigated over the past few years.[12–15]
There are two methods of PAI: photoacoustic tomography (PAT) and photoacoustic microscopy (PAM). In PAT, an array of transducers may be formed as linear, arc and circular shape, and mathematical algorithms are utilized for optical absorption reconstruction.[16,17] Also, for linear-array PAI, beamforming algorithms are needed to reconstruct the optical absorption distribution map of the tissue.[18–26] As the novel weighting methods, two modifications of coherence factor (CF) have been introduced.[27,28] Model-based (MB) algorithms build up a model to describe the relationship between detected PA signals and the reconstructed optical absorption distribution, iteratively reducing the artifacts.[29] Usually, in PAT, the acoustic detectors are considered point-like, and ignoring the size of the transducers leads to high frequencies information loss. MB algorithms can be used to compensate the lost of information by incorporating the detectors geometry in model matrix. It makes the imaging system computationally burdensome. Wavelet-packet framework can be used to reduce the model matrix size and make the use of inversion algorithms such as singular value decomposition (SVD) possible.[30] Efficient model of PAT image reconstruction based on discrete cosine transform (DCT) has been proposed in[31] and was enhanced using efficient block-sparse MB algorithm in which the number of required iterations was reduced by adopting a fast-converging optimization method.[32]
In PAT, limited number of transducers, especially when a human tissue in involved, leads to two main issues: high attenuation and low SNR. These drawbacks can be compensated using data averaging and a large number of elements of transducers enclosing the target of imaging. However, these solutions lead to a long data acquisition time and a large amount of data. Also, using a large number of elements of transducers leads to a high cost of imaging. So, there is a great value to tackle the problem and improve the image quality without increasing the hardware cost and data acquisition time.

In 2006, for the first time, Donoho *et al.* proposed compressed sensing (CS) theory which is based on a prior knowledge of unknowns.[33] This method can be used in the data acquisition procedure, which leads to information reconstruction from some observations that seem highly incomplete using a convex optimization.[34] CS theory has been used in Thermoacoustic Imaging (TAI).[35] In 2009, CS theory was used by Provost and Lesage in data acquisition procedure of PAT in which by taking advantage of sparse characteristic of samples, a small number of projections were used to reconstruct the optical absorption distribution.[36]

As illustrated in reference,[34] the answer of an optimization problem based on CS should have the minimum number of non-zero elements, defined as $\ell_0$-norm. Finding the minimum $\ell_0$-norm is an intractable problem because of the combinatorial search it imposes. Moreover, it is too sensitive to noise. As shown in,[37] $\ell_1$-norm can be used as problem constraint in which solution can be found by linear programming (LP) methods. In PAT, usually reconstruction algorithms based on CS use $\ell_1$-norm as constraint of optimization problem.[36] In this paper, it is proposed to use a smoothed version of $\ell_0$-norm ($S\ell_0$) to find the solution. It has been introduced in reference,[38] and it is shown that it provides a more accurate solution compared to basis pursuit. Here, we use $S\ell_0$ method to reconstruct the PA images based on the direct minimization of the $\ell_0$-norm which leads to a higher quality PA image.

## 2. MATERIALS AND METHODS

Considering a homogeneous and the thermal confinement, the recorded pressure $p(r,t)$ at position $r$ and time $t$ based on the PA effect can be written as follows:

$$\nabla^2 p(r,t) - \frac{1}{c^2}\frac{\partial^2}{\partial t^2}p(r,t) = -\frac{\beta}{C_p}\frac{\partial}{\partial t}H(r,t), \qquad (1)$$

where $\beta$ is the isobaric volume expansion coefficient, $H(r,t)$ is the heat sources, $c$ is the speed of sound, and $C_p$ is the heat capacity.[36] Assuming $H(r,t) = A(r)I(t)$, the forward problem of PAI can be written as follows:

$$p(r_0,t) = \frac{\beta}{4\pi C_p}\int \frac{dr'}{|r-r'|}\frac{\partial H(r',t')}{\partial t'}, \qquad (2)$$

where $t' = t - (|r-r'|/c)$. The analytical solution for (2) is reported for different geometries of imaging system.[39] Since the goal of this work is to integrate CS in image reconstruction procedure, a model is needed to make a connection between the acquired PA signals and the optical absorption distribution map of the tissue. The model can be written as follows:

$$y = Kx, \qquad (3)$$

where $y$ in the PA signals ($n \times 1$), $K$ ($n \times m$) is the imaging model, and $x$ ($m \times 1$) is the optical absorption (unknown of the model) which should be reconstructed. The problem of PA image reconstruction can be treated as an underdetermined linear equation. An underdetermined equation has a number answers.[40] An usual solution to address (3) is to choose the one that represents the smallest $\ell_2$ norm (among solutions in a range of $\epsilon$):

$$\min ||x||_{\ell_2} \; s.t. \; ||y - Kx||_{\ell_2} < \epsilon. \qquad (4)$$

In other word, minimization of the error term is the main idea of the solution introduced in (4). It can be represented as the minimization of the following formula:[36]

$$||y - Kx||_{\ell_2} + \mu ||x||_{\ell_2}. \qquad (5)$$

Mathematically, the minimization of the error does not provide the best solution. To find a more precise and accurate solution, the CS theory can be used. Lets assume that the image $x$ containing $N$ pixels to be reconstructed is sparse in a basis like $\phi$ where $\theta$ is the vector containing the coefficient of $x$ in the basis of $\phi$ ($x = \phi\theta$, $||x||_{\ell_0} < N$). In this case, the solution based on CS can be found using the following formula:

$$\min ||\theta||_{\ell_0} \; s.t. \; y = K\phi\theta. \qquad (6)$$

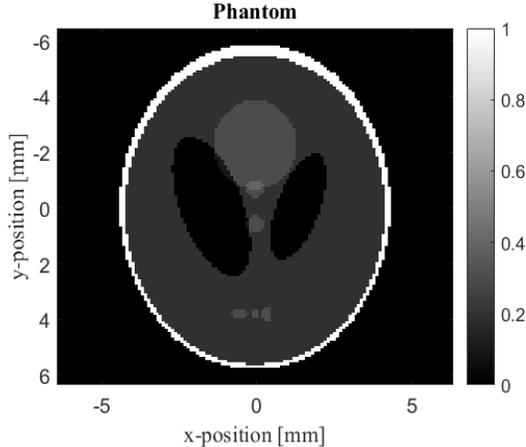
Figure 1: The phantom used in simulations.

Solving (6) is infeasible due to the need for a combinatorial search for its minimization and sensitivity to noise.[34] It has been proved that a more computationally efficient strategy for solving the convex problem in (6), it use the $\ell_1$ norm.[34] The problem can be written as follows:

$$\min||\theta||_{\ell_1} \ s.t. \ ||y - K\phi\theta||_{\ell_2} < \epsilon. \tag{7}$$

Contrary to all the previous methods used in PA image reconstruction (using $\ell_1$ norm minimization), in this paper, it is proposed to reconstruct the PA image using the direct minimization of the $\ell_0$ norm. We have used the S$\ell_0$ algorithm to solve the optimization problem mentioned in (6).[38] In this method, the discontinuous function of calculating $\ell_0$ is approximated by a suitable continuous one, which can be written as follows:

$$f_\sigma(s) = exp(\frac{-s^2}{2\sigma^2}), \tag{8}$$

where $\sigma$ determines the quality of the approximation. Considering

$$\lim_{\sigma \to 0} f_\sigma(s) = \begin{cases} 1, & s = 0 \\ 0, & s \neq 0 \end{cases}, \tag{9}$$

and, by defining

$$F_\sigma(s) = \sum_{i=1}^{m} f_\sigma(s_i), \tag{10}$$

the $\ell_0$-norm of a vector can be calculated using $||s||_{\ell_0} \approx m - F_\sigma(s)$. The evaluation of S$\ell_0$ is out of the scope of this paper. The readers are referred to reference[38] for further information. In the next section, simulation and results are explained to evaluate the performance of the proposed method.

## 3. RESULTS

To evaluate the proposed method, the wave propagation formula has been implemented. The imaging medium is 12.8 $mm$ × 12.8 $mm$ while 128 ×128 pixel is used (0.1 $mm$ for each side of a pixel). Sound speed is 1540 $m/s$. The Modified Shepp-Logan phantom (shown in Figure 1) has been used as the target of imaging. The sensors are positioned in the radius of 8 $mm$ with respect to the center of the phantom. 600 samples were recorded by each sensor, and a sampling frequency of 55 $MHz$ was used. The basis $\phi$ used for sparse representation of the signal is wavelet in this paper. We have compared the proposed method with iterative reconstruction (IR) and basis pursuit, and the results are shown in Figure 2. NOA represents the number of angles (sensors). As can be seen in Figure 2(a-c), IR algorithms is not able to follow the phantom of imaging well enough, even when NOA=64. On the other hand, basis pursuit results in a low quality image as a result of the low number of available samples for reconstruction. Based on the previous publications, the number of samples and sparsity of the imaging target highly affect the quality of reconstruction. In this paper, basis pursuit could not work well while the proposed algorithm (S$\ell_0$) shows a high performance. As can be seen in Figure 2(e), the formed image is highly similar to the imaging target (shown in Figure 1), and the proposed algorithm outperforms the other mentioned methods. The Figure 3 shows the lateral variations of the images shown in Figure 2. It can be seen that under the defined parameters for the simulation, the proposed method follows the phantom with a high precision while other methods, especially IR, do not provide a high precision in following the lateral variation of the phantom. We have used peak-signal-to-noise ratio (PSNR) metric to evaluate the results quantitatively. PSNR formula is as follows:

$$PSNR = 10.\log_{10}(\frac{N_x N_y}{\sum_{i=1}^{N_x}\sum_{j=1}^{N_y}(f_{i,j}-r_{i,j})^2}), \tag{11}$$

where $f$ is the gray-value of the reconstructed image, and r is the gray-value of the phantom. The size of the image is $Nx \times Ny$. PSNR for the images shown in Figure 2 are 18.51 $dB$, 5.13 $dB$, 22.7 $dB$, 18.79 $dB$ and 57.8 $dB$, respectively. The quantitative results show the superiority of the S$\ell_0$ while the number of the used sensors is lower than what used in the IR method. To put it more simply, S$\ell_0$ outperforms IR algorithm with a lower NOA and provides higher image quality compared to basis pursuit. To evaluate the performance of the proposed method, the number of the recorded samples is decreased to 500 (a sampling frequency of 45 $MHz$ ), and then the image reconstruction was applied. The results are shown in Figure 4. As can be seen, the quality of the reconstructed image using S$\ell_0$ is decreased compared to the image obtained with 600 samples (Figure 2(e)). However, it still outperforms the basis pursuit method.

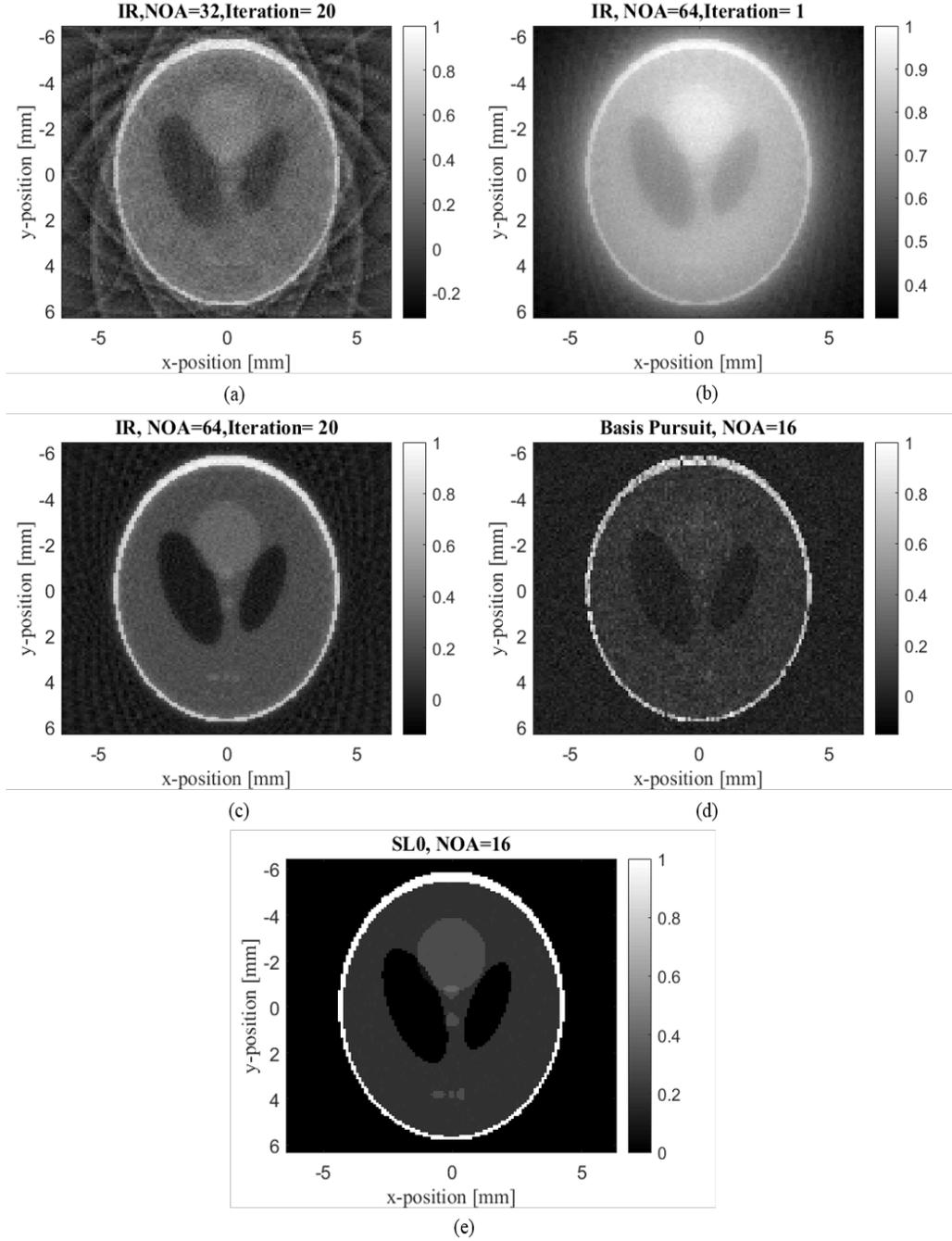

Figure 2: The reconstructed PA images using (a) IR algorithm with 20 iterations (NOA=32), (b) IR algorithm with 1 iteration (NOA=64), (c) IR algorithm with 20 iterations (NOA=64), (d) basis pursuit (NOA=16) and (e) S$\ell_0$ (NOA=16).

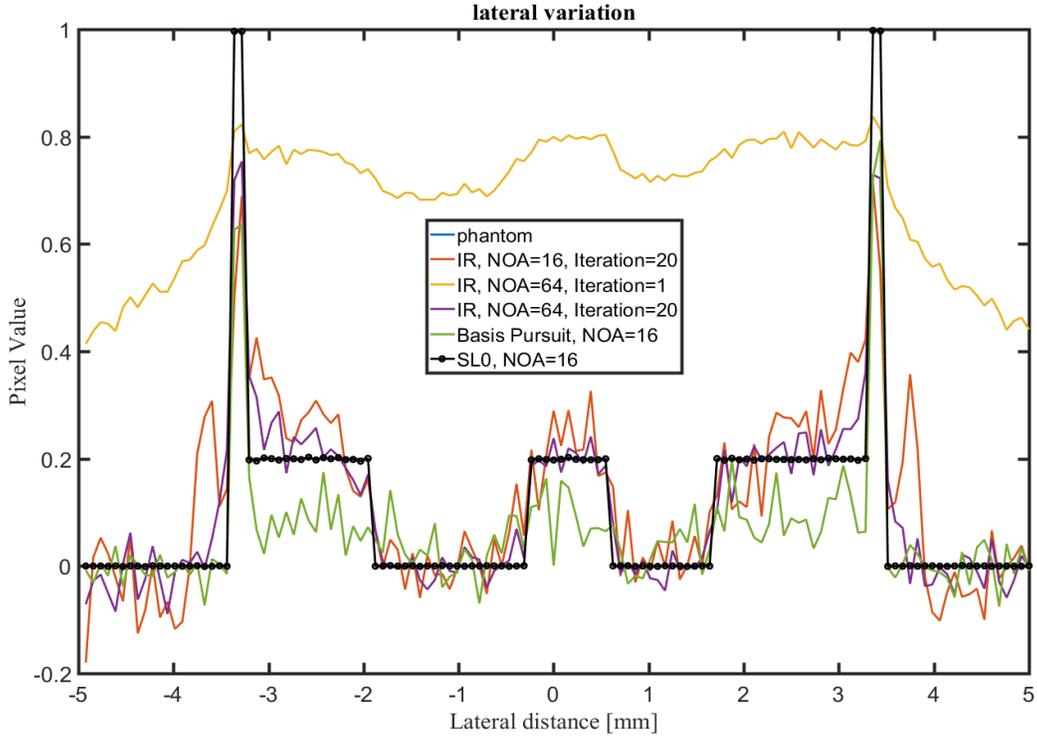

Figure 3: The lateral variations of the images shown in Figure 2.

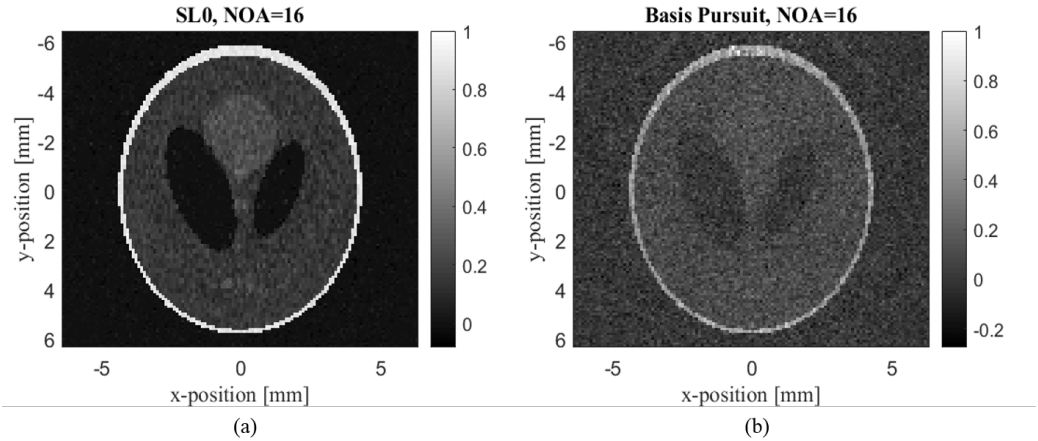

Figure 4: The reconstructed PA images using (a) S$\ell_0$ and (b) basis pursuit. NOA is 16, and the number of samples is 500.

## 4. CONCLUSION

While analytical PA image reconstruction algorithms provide a simple implementation and fast output, they suffer from a low accuracy and precision. To address these problems, MB algorithms are usually employed. However, a high number of sensors (angles) is needed to generate a high quality image. In this paper, we have used CS theory in MB image formation, and S$\ell_0$ was introduced as the reconstruction algorithm. The proposed method was compared to IR method and basis pursuit. It was shown that, considering the lateral variations and visual evaluation, S$\ell_0$ outperforms other methods, having only 16 sensors. The quantitative results showed that S$\ell_0$ improves the PSNR for about 39 $dB$, in comparison with basis pursuit, while all the imaging parameters were the same.

## REFERENCES

[1] Nasiriavanaki, M., Xia, J., Wan, H., Bauer, A. Q., Culver, J. P., and Wang, L. V., "High-resolution photoacoustic tomography of resting-state functional connectivity in the mouse brain," *Proceedings of the National Academy of Sciences* **111**(1), 21–26 (2014).
[2] Nasiriavanaki, M., Xia, J., Wan, H., Bauer, A. Q., Culver, J. P., and Wang, L. V., "Resting-state functional connectivity imaging of the mouse brain using photoacoustic tomography," in [*SPIE BiOS*], 89432O–89432O, International Society for Optics and Photonics (2014).
[3] Mahmoodkalayeh, S., Jooya, H. Z., Hariri, A., Zhou, Y., Xu, Q., Ansari, M. A., and Avanaki, M. R. N., "Low temperature-mediated enhancement of photoacoustic imaging depth," *arXiv preprint arXiv:1802.07114* (2018).
[4] Yao, J., Xia, J., Maslov, K. I., Nasiriavanaki, M., Tsytsarev, V., Demchenko, A. V., and Wang, L. V., "Noninvasive photoacoustic computed tomography of mouse brain metabolism in vivo," *Neuroimage* **64**, 257–266 (2013).
[5] Mehrmohammadi, M., Joon Yoon, S., Yeager, D., and Y Emelianov, S., "Photoacoustic imaging for cancer detection and staging," *Current molecular imaging* **2**(1), 89–105 (2013).
[6] Hariri, A., Fatima, A., and Nasiriavanaki, M., "A cost-effective functional connectivity photoacoustic tomography (fcpat) of the mouse brain," in [*Photons Plus Ultrasound: Imaging and Sensing 2017*], **10064**, 1006439, International Society for Optics and Photonics (2017).
[7] Khodaee, A. and Nasiriavanaki, M., "Study of data analysis methods in functional connectivity photoacoustic tomography (fcpat)," in [*Photons Plus Ultrasound: Imaging and Sensing 2017*], **10064**, 1006438, International Society for Optics and Photonics (2017).
[8] Mohammadi, L., Behnam, H., and Nasiriavanaki, M., "Modeling skull's acoustic attenuation and dispersion on photoacoustic signal," in [*Photons Plus Ultrasound: Imaging and Sensing 2017*], **10064**, 100643U, International Society for Optics and Photonics (2017).
[9] Xu, Q., Volinski, B., Hariri, A., Fatima, A., and Nasiriavanaki, M., "Effect of small and large animal skull bone on photoacoustic signal," in [*Photons Plus Ultrasound: Imaging and Sensing 2017*], **10064**, 100643S, International Society for Optics and Photonics (2017).
[10] Volinski, B., Hariri, A., Fatima, A., Xu, Q., and Nasiriavanaki, M., "Photoacoustic investigation of a neonatal skull phantom," in [*Photons Plus Ultrasound: Imaging and Sensing 2017*], **10064**, 100643T, International Society for Optics and Photonics (2017).

[11] Meimani, N., Abani, N., Gelovani, J., and Avanaki, M. R., "A numerical analysis of a semi-dry coupling configuration in photoacoustic computed tomography for infant brain imaging," *Photoacoustics* **7**, 27–35 (2017).

[12] Hariri, A., Fatima, A., Mohammadian, N., Mahmoodkalayeh, S., Ansari, M. A., Bely, N., and Avanaki, M. R., "Development of low-cost photoacoustic imaging systems using very low-energy pulsed laser diodes," *Journal of biomedical optics* **22**(7), 075001 (2017).

[13] Hariri, A., Bely, N., Chen, C., and Nasiriavanaki, M., "Towards ultrahigh resting-state functional connectivity in the mouse brain using photoacoustic microscopy," in [*SPIE BiOS*], 97085A–97085A, International Society for Optics and Photonics (2016).

[14] Nasiriavanaki, M. et al., "Resting-state functional connectivity measurement in the mouse brain using a low cost photoacoustic computed tomography," in [*Frontiers in Optics*], JW4A–62, Optical Society of America (2016).

[15] Hariri, A., Hosseinzadeh, M., Noei, S., and Nasiriavanaki, M., "Photoacoustic signal enhancement: towards utilization of very low-cost laser diodes in photoacoustic imaging," in [*Photons Plus Ultrasound: Imaging and Sensing 2017*], **10064**, 100645L, International Society for Optics and Photonics (2017).

[16] Xu, M. and Wang, L. V., "Time-domain reconstruction for thermoacoustic tomography in a spherical geometry," *IEEE transactions on medical imaging* **21**(7), 814–822 (2002).

[17] Omidi, P., Diop, M., Carson, J., and Nasiriavanaki, M., "Improvement of resolution in full-view linear-array photoacoustic computed tomography using a novel adaptive weighting method," in [*Photons Plus Ultrasound: Imaging and Sensing 2017*], **10064**, 100643H, International Society for Optics and Photonics (2017).

[18] Mozaffarzadeh, M., Mahloojifar, A., and Orooji, M., "Medical photoacoustic beamforming using minimum variance-based delay multiply and sum," in [*Digital Optical Technologies 2017*], **10335**, 1033522, International Society for Optics and Photonics (2017).

[19] Mozaffarzadeh, M., Mahloojifar, A., and Orooji, M., "Image enhancement and noise reduction using modified delay-multiply-and-sum beamformer: Application to medical photoacoustic imaging," in [*Iranian Conference on Electrical Engineering (ICEE) 2017*], 65–69, IEEE (2017).

[20] Mozaffarzadeh, M., Avanji, S. A. O. I., Mahloojifar, A., and Orooji, M., "Photoacoustic imaging using combination of eigenspace-based minimum variance and delay-multiply-and-sum beamformers: Simulation study," *arXiv preprint arXiv:1709.06523* (2017).

[21] Mozaffarzadeh, M., Mahloojifar, A., Orooji, M., Kratkiewicz, K., Adabi, S., and Nasiriavanaki, M., "Linear-array photoacoustic imaging using minimum variance-based delay multiply and sum adaptive beamforming algorithm," *Journal of Biomedical Optics* **23**(2), 026002 (2018).

[22] Mozaffarzadeh, M., Mahloojifar, A., Nasiriavanaki, M., and Orooji, M., "Eigenspace-based minimum variance adaptive beamformer combined with delay multiply and sum: Experimental study," *arXiv preprint arXiv:1710.01767* (2017).

[23] Paridar, R., Mozaffarzadeh, M., Mehrmohammadi, M., and Orooji, M., "Photoacoustic image formation based on sparse regularization of minimum variance beamformer," *arXiv preprint arXiv:1802.03724* (2018).

[24] Mozaffarzadeh, M., Mahloojifar, A., Orooji, M., Adabi, S., and Nasiriavanaki, M., "Double-stage delay multiply and sum beamforming algorithm: Application to linear-array photoacoustic imaging," *IEEE Transactions on Biomedical Engineering* **65**(1), 31–42 (2018).

[25] Mozaffarzadeh, M., Sadeghi, M., Mahloojifar, A., and Orooji, M., "Double-stage delay multiply and sum beamforming algorithm applied to ultrasound medical imaging," *Ultrasound in Medicine and Biology* **44**(3), 677 – 686 (2018).

[26] Paridar, R., Mozaffarzadeh, M., Nasiriavanaki, M., and Orooji, M., "Double minimum variance beamforming method to enhance photoacoustic imaging," *arXiv preprint arXiv:1802.03720* (2018).

[27] Mozaffarzadeh, M., Yan, Y., Mehrmohammadi, M., and Makkiabadi, B., "Enhanced linear-array photoacoustic beamforming using modified coherence factor," *Journal of Biomedical Optics* **23**(2), 026005 (2018).

[28] Mozaffarzadeh, M., Mehrmohammadi, M., and Makkiabadi, B., "Image improvement in linear-array photoacoustic imaging using high resolution coherence factor weighting technique," *arXiv preprint arXiv:1710.02751* (2017).

[29] Zhang, C., Zhang, Y., and Wang, Y., "A photoacoustic image reconstruction method using total variation and nonconvex optimization," *Biomedical engineering online* **13**(1), 1 (2014).

[30] Rosenthal, A., Jetzfellner, T., Razansky, D., and Ntziachristos, V., "Efficient framework for model-based tomographic image reconstruction using wavelet packets," *IEEE transactions on medical imaging* **31**(7), 1346–1357 (2012).

[31] Zhang, Y., Wang, Y., and Zhang, C., "Efficient discrete cosine transform model–based algorithm for photoacoustic image reconstruction," *Journal of biomedical optics* **18**(6), 066008–066008 (2013).

[32] Zhang, C., Wang, Y., and Wang, J., "Efficient block-sparse model-based algorithm for photoacoustic image reconstruction," *Biomedical Signal Processing and Control* **26**, 11–22 (2016).

[33] Donoho, D. L., "Compressed sensing," *IEEE Transactions on information theory* **52**(4), 1289–1306 (2006).

[34] Candès, E. J., Romberg, J., and Tao, T., "Robust uncertainty principles: Exact signal reconstruction from highly incomplete frequency information," *IEEE Transactions on information theory* **52**(2), 489–509 (2006).

[35] Zhu, X., Zhao, Z., Wang, J., Song, J., and Liu, Q. H., "Microwave-induced thermal acoustic tomography for breast tumor based on compressive sensing," *IEEE Transactions on Biomedical Engineering* **60**(5), 1298–1307 (2013).

[36] Provost, J. and Lesage, F., "The application of compressed sensing for photo-acoustic tomography," *IEEE transactions on medical imaging* **28**(4), 585–594 (2009).

[37] Donoho, D. L., Elad, M., and Temlyakov, V. N., "Stable recovery of sparse overcomplete representations in the presence of noise," *IEEE Transactions on information theory* **52**(1), 6–18 (2006).

[38] Mohimani, H., Babaie-Zadeh, M., and Jutten, C., "A fast approach for overcomplete sparse decomposition based on smoothed l0 norm," *IEEE Transactions on Signal Processing* **57**(1), 289–301 (2009).

[39] Xu, M., Xu, Y., and Wang, L. V., "Time-domain reconstruction algorithms and numerical simulations for thermoacoustic tomography in various geometries," *IEEE Transactions on biomedical engineering* **50**(9), 1086–1099 (2003).

[40] Hayashi, K., Nagahara, M., and Tanaka, T., "A user's guide to compressed sensing for communications systems," *IEICE transactions on communications* **96**(3), 685–712 (2013).